\documentstyle[aps,twocolumn,epsf,rotate,prl]{revtex}
\newcommand{\be}{\begin{equation}}
\newcommand{\ee}{\end{equation}}
\newcommand{\bdm}{\begin{displaymath}}
\newcommand{\edm}{\end{displaymath}}
\newcommand{\ba}{\begin{eqnarray}}
\newcommand{\ea}{\end{eqnarray}}

\begin{document}

%\draft
%\twocolumn[\hsize\textwidth\columnwidth\hsize\csname @twocolumnfalse\endcsname

\title{
Internal Vortex Structure of a Trapped Spinor Bose-Einstein Condensate 
} 
\author{S.-K. Yip}
\address{ 
Physics Division, National Center for Theoretical Sciences ,
P. O. Box 2-131,  Hsinchu, Taiwan 300, 
R. O. C.} 
\date{\today}

\maketitle

\begin{abstract}
{\small

The internal vortex structure of a 
trapped spin-$1$ Bose-Einstein condensate is investigated.
It is shown that it has a variety of configurations
depending on, in particular, the ratio of the relevant
scattering lengths and the total magnetization.

 }

\end{abstract}
\date{\today}
\vspace*{0.2 cm}

%\begin{multicols}{2}

Recently the MIT group has succeeded in obtaining Bose Einstein
Condensation (BEC)
of $^{23}$Na atoms in an optical trap. \cite{Stamper,Stenger}
A novel aspect of this system is that $^{23}$Na atoms
possess a hyperfine spin, with $f =1$ in the lower multiplet.
All three possible projections of the hyperfine spin can be
optically trapped simultaneously.  Thus generally the condensate
have to be described by a spin-$1$ order parameter consisting
of the spinor $(\Psi_u,\Psi_0,\Psi_d)$ where $\Psi_{u,0,d}$
are the macroscopic wavefunctions with the hyperfine projection 
$s = 1, 0, -1$ respectively. 
%  This is
%akin to the case of $^{3}$He \cite{Vollhardt} and 
%some heavy fermion systems \cite{Heffner96},
%where the superfluid instabilities are also towards 
%multicomponent order parameters. 
%There for temperatures
%below the critical, the symmetry within the order parameter
%space is spontaneously broken, with the state determined
% by minimization of the quartic term in the
%Ginzburg-Landau free energy.  The analogous problem
%for the $^{23}$Na condensate has been worked out
%by Ho\cite{Ho98} and independently Ohmi and Machida \cite{Ohmi98}.
%In the Bose-condensed phase the spin rotational symmetry is
%spontaneously broken with
The ground state of the condensate is determined by
the spin-dependence of the interaction between the bosons.
\cite{Ho98,Ohmi98}
In the dilute limit they can be characterized by the (s-wave) 
scattering lengths $a_0$ and $a_2$ in the total (hyperfine-)spin
$0$ and $2$ channels respectively.
The polar state, where the order parameter is
proportional to the spinor $(0, 1, 0)$,  is favored if $a_2 > a_0$, but the
axial state, where the spinor is $(1, 0, 0)$, if the inequality 
is otherwise. $^{23}$Na belongs to the former case.
However, $^{87}$Rb is predicted to belong to the latter case.
\cite{Ho98} 

A particular interesting feature of these systems  is that,
due to the weak spin dependence of the interatomic interaction,
 $a_0 \approx a_2$.  For example  for $^{23}$Na
%ref \cite{Ho98} quoted
$a_0 \approx 46 a_B$ and $a_2 \approx 52 a_B$, where $a_B$ is the Bohr
radius.  Thus the difference $a_2- a_0 \approx 6 a_B  $ is only a
small fraction of $a_0$ or $a_2$.  As a result,
the condensate only needs to pay very little extra energy
to get into the ``wrong'' state.  
If there is a competing energy, such as that due to
the presence of a gradient, it may be energetically 
favorable for the condensate to deviate locally from
the polar state.
An analogous remark
is also applicable to $^{87}$Rb, where 
%according to
%numbers cited by ref \cite{Ho98}, 
$a_0 \approx 110 a_B$ and
$a_2 \approx 107a_B$. \cite{Burke}

In this paper I shall illustrate this by considering
 vortices of  this spin-$1$
Bose condensate. 
%While
%there are no successful reports of producing/observing
%vortices so far, since the existence of quantized vortices is a unique
%property of quantum superfluids, an observation
%of these objects would be interesting indeed.  
Vortices
with a scalar order parameter in BEC have been discussed
in many papers (e.g \cite{Dodd97,Rokhsar97} ). 
In a cylindrically symmetric trap, 
there is no stable vortex state for angular momentum 
 $L < N \hbar$ where $N$ is the number of particles.
A singly quantized vortex has $L = N \hbar$, 
%the vortex is located at the center of the trap
with the node of the order parameter 
located at the center of the trap.
 Here I shall show that
vortices in a spinor condensate are even more interesting
in that they exhibit a very rich internal structure.
In general locally the order parameter is
in neither the polar nor the axial state.
They may have broken cylindrical symmetry with nodes of
the order parameter of
individual species appearing at positions other than
the trap center.  The minimum angular momentum required
for the formation of a vortex is also less than $N$.
Moreover  transitions between
different internal vortex structures are possible.
Vortices were also discussed in Ref \cite{Ho98} and \cite{Ohmi98},
but they did not consider structures
which arise from deviations of the
order parameter from the original polar or axial phases.

Since the order parameter has more than one component,
it is convenient to distinguish between vortices of individual
order parameter component $\Psi_s$ and the composite
structure.  
I shall refer the latter as the composite vortex
(CV).
%I shall follow ref \cite{Tokuyasu}
%and call the latter Vortices.  
%I shall also
%refer to the vortex with a scalar (single component)
%order parameter a scalar vortex. (SV)

The order parameter, in particular that
of the CV, is found by minimization of the energy
(restricting ourselves to $T = 0$) under appropriate constraints.
The energy density ${\cal E}$ consists of 
the kinetic and potential contributions 
$ \sum_s { | \nabla \Psi_s |^2 \over 2 M_a} + V | \Psi_s |^2$,
where $M_a$ is the atomic mass,
$V$ is the trap potential and the sum is over all
spin components,
and the interaction part which can be written as
$ {\cal E}_{\rm int}  = { 1 \over 2} (c_0 + c_2) n^2 
  - { 1 \over 2} c_2 | 2 \Psi_u \Psi_d - \Psi_0|^2 $
where $n = \sum_s |\Psi_s|^2 $ is the local density.
Here $c_0 \equiv { g_0 + 2 g_2 \over 3 }$ and 
$ c_2 \equiv {g_2 - g_0 \over 3}$
where $g_{0,2}$ are in turn related to the scattering
lengths in the total spin $0$ and $2$ channels
via $g_{0,2} = { 4 \pi \hbar^2 a_{0,2} \over M_a}$. \cite{Ho98}
The total particle number $N$,
angular momentum $L$ and magnetization $M$ should
be considered as conserved if no exchange of the corresponding quantity is
allowed between the atoms inside the trap and
their environment (within the relevant experimental time scale).
In this case we have to minimize the total energy
for given $N$, $L$ and $M$.  As usual
it is convenient to introduce and minimize the free energy
$ G \equiv E - \mu N - \Omega L - H M$
%$F = \int ( {\cal E} - \mu n - \Omega {\cal L} - h m_z )$
where  $\mu, \Omega$ and $H$
are Lagrange multipliers. 
%to produce the desired $N$, $L$ and $M$.
$\mu, \Omega, H$  correspond to the
chemical potential, angular velocity 
and magnetic field.
% in the language of condensed matter physics.
$\Omega$ is given by the angular
velocity of the rotating trap if angular momentum
can be exchanged between the trapped atoms and their environment.
%${\cal L} \equiv {\cal L}_s =
%\sum_s \vec r \times { \hbar \over 2 i}
% [\Psi^{*}_s \nabla \Psi_s
% - (\nabla \Psi^{*}_s) \nabla \Psi_s ] $
%is the angular momentum density,
%$m_z =  \Psi^{*}{}_u \Psi_u - \Psi^{*}{}_d \Psi_d $
%is the magnetization.
%(need $m_{x,y}$?)

It is useful to note that the energy is 
invariant under relative rotation between the
real and spin space. Accordingly
in below the spin quantization axis will be
chosen for the most convenient presentation 
(and always along the total magnetization if
it is finite).  In particular the configurations
presented below do not rely on any special
relative orientation between the net magnetization
and the rotational axis (which is always chosen as $z$).
%Since the transformation of
%a vector under rotation is simplier than that of a 
%Since the rotational properties of the magnetization
%are simplier than those of the
%spin-$1$ wavefunction.
 I shall also discuss
the local magnetization density $\vec m$. The projection of 
$\vec m$  along 
a general direction is measureable in BEC experiments
since it is given by the difference in the number
density between the $u$ and $d$ species 
when one uses that direction as the quantization axis. 

Setting the variation of the free energy with respect
to $\Psi_s^{*}$ to zero,
 one obtains the familiar Gross-Pitaevskii (GP)
equations (generalized due to the presence of multiple spin species
\cite{Ho98,Ohmi98}).
If $c_2 > 0$ such as in the case of $^{23}$Na,  in
the absence of a net magnetization the order parameter
 can be {\it chosen} so that only
$\Psi_0$ is finite and obeys the GP
equation in the usual form:
$0$ $=$ $( -  { \hbar^2 \over 2 M_a} \nabla^2 + V - \mu ) \Psi_0$
   $ +$$ c_0 | \Psi_0|^2 \Psi_0$.
The order parameter profile for $ \Psi_0$ would then be 
completely analogous to that of a scalar order parameter
with the interaction parameter given by $c_0$ 
(note then $\Psi_0$ is independent of $c_2$ ).
In particular in the absence of any circulation
and if one ignores 
the gradient term,
( the Thomas Fermi (TF) approximation \cite{Baym96})
$ | \Psi_0 |^2 = { \mu - V \over c_0} \theta ( \mu - V)$.
The structure of a singly quantized vortex would also
be exactly analogous to that of a scalar order parameter
 investigated by, {\it e.g.},  Dodd et al \cite{Dodd97}.
%The vortex is located at the center of the trap,
%with $L/N = 1$. 
  For the discussions below it is also convenient
to re-consider 
the same CV with quantization axis rotated 
by $\pi/2$ about a horizontal axis.  In this basis
the above CV becomes two coinciding vortices
of the $u$ and $d$ components with $|\Psi_u| = |\Psi_d|$ 
and their nodes at the trap center.
We shall see below that in general the CV
is very different from the ones just discussed.
%The CV may have broken cylindrical symmetry with vortices of 
%individual species appearing at positions other than
%the trap center.  The minimum angular momentum required
%for the formation of the CV is also less than $N$.

Typically in the experiments the cloud is trapped
by a potential harmonic in all three spatial directions.
For numerical simplicity I shall instead consider a cloud 
subject to a harmonic potential only in the $x-y$ plane
but of uniform density within thickness $d$ along the rotational $z$
axis.
It is
reasonable to assume that the results below will be qualitatively
applicable to a pancake shaped cloud trapped by a three dimensional,
 axially symmetric  harmonic potential if the 
radii of the clouds perpendicular to the rotational axis are comparable.
Rather than varying $\mu$ and $\Omega$ to obtain a fixed total
number of particles and angular momentum,
 I shall simply present the types of CV
% (the "phase diagram")
for fixed $\mu$'s and $\Omega$'s. 
However, I shall continue to use the total magnetization
(rather than $H$) as an independent variable \cite{Stenger}.
 I shall eliminate $\mu$ in favor
of $R \equiv ( 2 \mu/ M_a \omega_o^2)^{1/2}$,
%here $\alpha$ is the curvature of the harmonic trapping potential.
%($ V = { 1 \over 2 } \alpha r^2$) 
where $\omega_o$ is the (angular) trap frequency.
In the absence of vortices and under the TF approximation, 
the radius of the cloud  
and the total number of particles  are 
independent of the value of $c_2$ and given by $R$ and 
$ N_o = { d \over 16 a }( { R \over \lambda_o} )^4$ respectively,
where $\lambda_0$ is the
size of the harmonic oscillator ground state wavefunction
($\lambda_o = ( \hbar /M_a \omega_o)^{1/2}$).
Here 
%$d$ is the thickness of the cloud,
$a \equiv {(a_0 + 2 a_2) \over 3 }$ is an effective scattering length
for the interaction parameter $c_0$.
All the CV presented below has $N \approx N_o$.
\cite{noteN}

\begin{figure}[h]
\centerline
{ \epsfxsize=0.27\textwidth
\epsfysize=0.37 \textwidth
\rotate[r]
{ \epsfbox{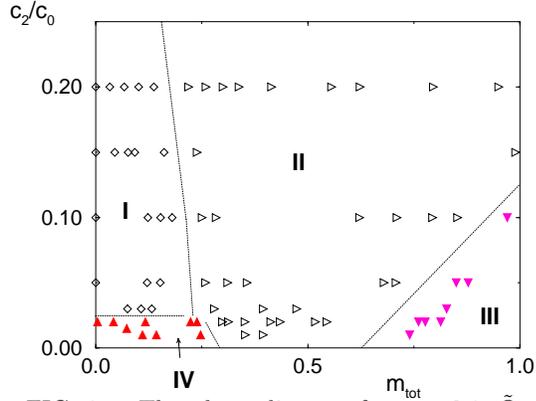} }
}
\vskip 0.4  cm
\begin{minipage}{0.45\textwidth}
\caption[]{
The phase diagram for $\epsilon = 0.1$, $\tilde \Omega = 0.45$.
I (diamonds):
vortices in both $u$ and $d$,
configuration as in Fig \ref{fig:rg0.2dx2}
except for a possible adjustment of relative magnitudes
of $|\Psi_{u,d}|$;
II  (triangles right): vortex in $u$ with $d$ core as in
Fig \ref{fig:rg0.2hx2};
III (triangles down): vortex in $u$ with mainly $0$ core
as in Fig \ref{fig:rg0.02h_psi};
IV (triangles up): configuration as in Fig. \ref{fig:rg0.02d_psi}.
Dotted lines are guides to the eye.
}

%\vskip 0.3 cm
\label{fig:pd}
\end{minipage}
\end{figure}

I shall introduce the parameter $\epsilon \equiv
( \lambda_0 /R)^2$ which measures the
deviation from the TF ($\epsilon \rightarrow 0$) limit.
$\epsilon$ depends only weakly on $N$
for given trap parameters.
I shall
 express $\Psi_s$ in units of $\sqrt{\mu/c_o}$
(correspondingly the number density $n$ and the magnetization
density $\vec m$ in $\mu/c_o$) 
%\cite{formulas},
distances in units of $R$, 
%energies and free energies
%in units of $F_o \equiv {\pi \over 4} { (\hbar \omega_o)^2 \over c_o}
%{R^4 \over \lambda_o^2} d $, 
and total particle number and magnetization
 $m_{\rm tot}$ in units of $N_o$.  
With this, all physical results depend only on the dimensionless
parameters $\epsilon$, $\tilde \Omega \equiv \Omega/\omega_o$,
$\tilde c_2 \equiv c_2/c_o$ and $m_{\rm tot}$.
Anticipating future experiments on other atoms I will
not fix $\tilde c_2$ to that of $^{23}$Na
(though confining myself to $\tilde c_2 > 0$ ).
 As a concrete example I shall consider
mainly $\epsilon = 0.1$, $\tilde \Omega = 0.45 $,
\cite{omega}
with the corresponding phase diagram  shown in Fig. \ref{fig:pd}.
I shall comment on other values of the parameters
as I proceed.

\begin{figure}[h]
\centerline
{ \epsfxsize=0.22\textwidth
\epsfysize = 0.37\textwidth
\rotate[r]
{ \epsfbox{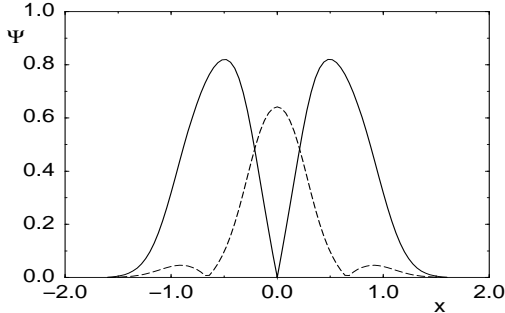} } }
\vskip 0.3 cm
\begin{minipage}{0.45\textwidth}
\caption[]{ CV for $\tilde c_2 = 0.2$, $\tilde \Omega = 0.45$.
$m_{\rm tot} = 0$. $|\Psi_0|$, full line;
 $|\Psi_u| = |\Psi_d|$, dashed.
}

%\vskip 0.3 cm
\label{fig:rg0.2dz1}
\end{minipage}
\end{figure}

We begin by considering 
 $m_{\rm tot} = 0$.
I shall present the CV in two ways,
each related to some of
the CV structures discussed for 
$m_{\rm tot} \ne 0$ below.
The structure of the CV with quantization axis
chosen so that it resembles  most closely
 a vortex of $\Psi_0$ alone
is as shown in Fig. \ref{fig:rg0.2dz1}.
%The CV is not simply a vortex of $\Psi_0$ alone.
However, instead of an empty core,
it is energetically favorable for some of the
$0$ particles to convert to $u$ and $d$ species
and  appear near the center of the trap.
%[this is supported by the fact that the (dimensionless) free energy
%of this CV ($-1.505$) is less than that of the SV
%($-1.46$)]
For the present parameters, $|\Psi_{u}| = |\Psi_{d}|$ 
and each has two nodes with unit circulation.
In Fig \ref{fig:rg0.2dz1} the order parameter
along the line (chosen as the $x$ axis) going through these singularities 
was shown.  Note that the CV has broken cylindrical 
symmetry. \cite{Chui}
% The gauge can be chosen
%so that $\Psi_0$ ($\Psi_{u}=\Psi_{d}$ )
% is real and positive for $x > 0$ ($x \approx 0$).
%Since $\Psi_{u}=\Psi_{d}$, $m_y = m_z = 0$ everywhere.
%However, locally $m_x$ is finite. 
%[Not shown, but qualitatively it behaves similar to that of
%Fig \ref{fig:rg0.02h_m} below, see also next paragraph.]

%%%%%%%%%%%%%%%%%%%%%%%%%%%%%%%%%%%%%%%%%%%%%%%%%%%%%%%%%%%

%For larger $\tilde c_2$ (more tendency towards polar),
%the $\Psi_{u,d}$ magnitude decreases
%and vortices come
%closer together.  Correspondingly the magnitude
%of $m_x$ and the region where it is finite
%move closer to the trap center.  However, the 
%vortex is still distinguishable from a scalar
%vortex consisting of $\Psi_0$ alone even when
%$\tilde c_2 \approx 1.0$.

\begin{figure}[h]
\centerline
{ \epsfxsize=0.22\textwidth
\epsfysize = 0.37 \textwidth
\rotate[r]
{ \epsfbox{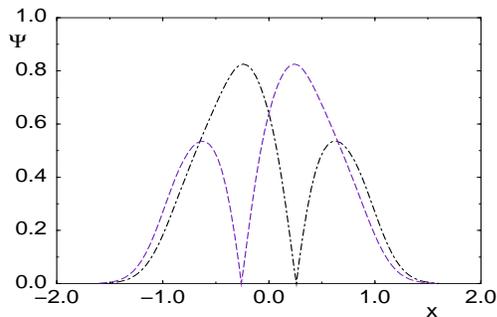} } }
\vskip 0.3 cm
\begin{minipage}{0.45\textwidth}
\caption[]{ The same CV as Fig \ref{fig:rg0.2dz1},
but with different quantization axis.
%Upper panel: $\Psi_0$: full line, $\Psi_u = \Psi_d$, dashed line.
 $|\Psi_u|$, dashed-dotted; $|\Psi_d|$, dashed.
}

\label{fig:rg0.2dx2}
\end{minipage}
\end{figure}

Another useful
way of presenting the above CV
is to use quantization axis rotated by $\pi/2$
about a horizontal axis with respect to those above.
In this basis only $\Psi_{u}$ and $\Psi_{d}$
are finite. Each has one node, displaced
by equal but opposite distance from the trap center  
(Fig \ref{fig:rg0.2dx2}).  It follows that
the local magnetization density is finite and points
along the (present) $\hat z$-axis, being
negative for $x< 0$ and positive for $x>0$.  
Notice that at the singularity for say the $d$-component,
since $|\Psi_d| = 0$ and $|\Psi_u| \ne 0$, locally
the condensate is actually in the axial but
not polar state (even though $m_{\rm tot}$ and $H$ are zero)

 With the use of this quantization axis we can also
understand easily the reason for the present CV structure.
Due to
the presence of the trap potential a vortex has maximum kinetic
energy if its node is located at the center of the trap.
It is thus energetically favorable for the nodes of 
the $u$ and $d$ components to move
away from the trap center.  \cite{Dodd97,Rokhsar97}
 For the system to be at an energy minimum,
they 
move opposite to each other,
creating regions where $|\Psi_u| \ne |\Psi_d|$,
%The tendency of the vortices to move apart due to
%the trap potential  is
eventually balanced by the desire 
of the condensate to remain in the polar state. 
In this picture it is obvious that 
 $L /N < 1$.
As $\Omega$ increases, the $\Psi_{u,d}$ singularities
move closer to the center of the trap and $L/N$ increases.
[{\it e.g.}, for $\tilde c_2 = 0.2$, $L/N = 0.87 (0.92)$ at
 $\tilde \Omega = 0.45 (0.5)$].
The separation
between these singularities, and the region where 
the local magnetization is non-zero, increase
with decreasing $\tilde c_2$  [correspondingly 
$L/N$ decreases: {\it e.g.}, at $\tilde \Omega = 0.45$, 
 $L/N = 0.84 (0.80)$ for
$\tilde c_2 = 0.1 (0.05)$]

%%%%%%%%%%%%%%%%%%%%%%%%%%%%%%%%%%%%%%%%%%%%%%%%%%%%%%%%%%%

Now we are ready to consider the structure of the CV with finite
total magnetization.  I shall describe
each region of the phase diagram Fig. \ref{fig:pd} in turn.

{\bf I:} In this region the favorable configuration is similar to
that of Fig \ref{fig:rg0.2dx2} except for an increase (decrease) in
the amplitude of $\Psi_u$ and $\Psi_d$ (not shown).
The local and total magnetization of this CV
are always collinear.
One can understand this configuration by
considering the energy under the magnetic field $H$.
  The CV has an order parameter and
hence a magnetic susceptibility which 
is anisotropic.  
For a given magnitude of the magnetization, 
the energy is minimum if the direction 
of $\vec m_{\rm tot}$  is along that of largest
susceptibility.  
The quantization axes used
in Fig \ref{fig:rg0.2dz1} and \ref{fig:rg0.2dx2} above
correspond to 
the principal directions of the susceptibility tensor.
It is intuitively reasonable that the CV 
has larger susceptibility along the quantization axis
of Fig \ref{fig:rg0.2dx2} ({\it c.f.} \cite{Ohmi98}). 

%%%%%%%%%%%%%%%%%%%%%%%%%%%%%%%%%%%%%%%%%%%%%%%%%%%%%%%%%%%%%%%%%

\begin{figure}[h]
\centerline
{ \epsfxsize=0.22\textwidth
\epsfysize=0.37\textwidth
\rotate[r]
{ \epsfbox{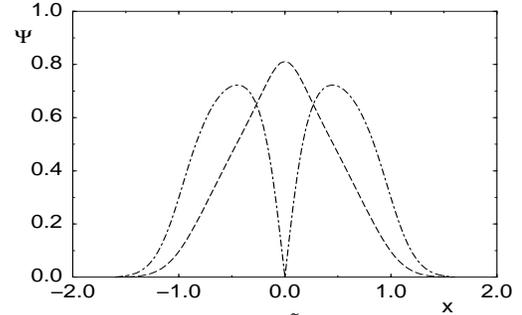} } }
\vskip 0.3 cm
\begin{minipage}{0.45\textwidth}
\caption[]{ CV for $\tilde c_2 = 0.2$, $\tilde \Omega = 0.45$,
and $m_{\rm tot} =0.414 $. 
%($H =0.04 $).
 $|\Psi_u|$, dashed-dotted; $|\Psi_d|$, dashed.
}

\label{fig:rg0.2hx2}
\end{minipage}
\end{figure}

{\bf II: }
For larger $m_{\rm tot}$ 
the vortex of
the minority species $d$ disappears.
  The CV is replaced by a vortex of the $u$ species
with a $d$ core.
( Fig. \ref{fig:rg0.2hx2}.)  This can be understood by
recognizing that the effective chemical potential
for the $d$ species is given by $\mu - H$.  Increasing
 $m_{\rm tot}$ requires increasing $H$, hence decreasing
$\mu - H$.  Eventually the effective chemical potential
is too low to overcome the necessary kinetic energy 
required for forming a circulating $d$ component.
This picture is supported by the fact that the
critical $m_{\rm tot}$ needed for the I $\rightarrow$ II transition
increases with $\Omega$.  The $- \Omega L$ term
in the free energy favors an order parameter
with finite circulation, thus at higher angular velocity
a larger $H$ and hence $m_{\rm tot}$ is required for the transition.
[{\it e.g},  at $\tilde c_2 = 0.2$, 
 the critical $m_{\rm tot} \approx 0.2$ for $\tilde \Omega = 0.45$ 
here (Fig \ref{fig:pd}) whereas $m_{\rm tot}  \approx 0.4$ 
for $\tilde \Omega  = 0.5$]

%%%%%%%%%%%%%%%%%%%%%%%%%%%%%%%%%%%%%%%%%%%%%%%%%%%%%%%%%%%%%%%%%%

\begin{figure}[h]
\centerline
{ \epsfxsize=0.22\textwidth
\epsfysize=0.37\textwidth
\rotate[r]
{ \epsfbox{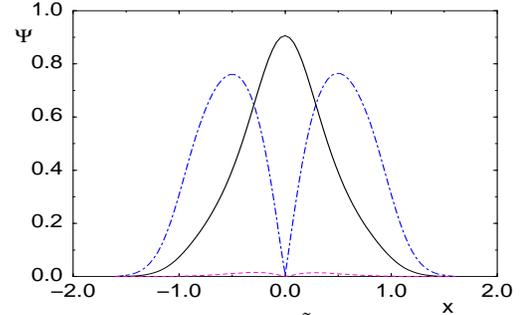} } }
\vskip 0.3 cm
\begin{minipage}{0.45\textwidth}
\caption[]{ CV for $\tilde c_2 = 0.02$, $\tilde \Omega = 0.45$,
and $m_{\rm tot} =0.77 $. 
%($H =0.015 $).
 $|\Psi_u|$, dashed-dotted; $|\Psi_0|$, full-line.
}

%\vskip 0.3 cm
\label{fig:rg0.02h_psi}
\end{minipage}
\end{figure}

{\bf III:} This occurs at still larger $m_{\rm tot}$
and only for small $c_2$.
In this region the CV has a u vortex
 with a core filled by the $0$ species (Fig \ref{fig:rg0.02h_psi}).
The minimum magnitude of $m_{\rm tot}$ needed for
this new CV increases with $\tilde c_2$.
These features 
 can be understood by considering again the effective
chemical potential for the $0$ and $d$ spins which
are $\mu$ and $\mu-H$ respectively.
 The spin $0$ species is more favored by $H$,
but suffers a stronger repulsion (than
the $d$ species) from the majority $u$ species due
to the spin dependent interaction $c_2 (>0)$.
Only at sufficiently small $c_2$ and large $m_{\rm tot}$
does this CV
become favorable.

\begin{figure}[h]
\centerline
{ \epsfxsize=0.22\textwidth
\epsfysize=0.37\textwidth
\rotate[r]
{ \epsfbox{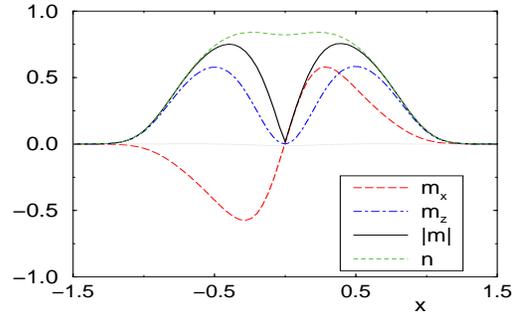} } }
\vskip 0.3 cm
\begin{minipage}{0.45\textwidth}
\caption[]{ Local magnetization for the CV of Fig \ref{fig:rg0.02h_psi}.
 }

%\vskip 0.3 cm
\label{fig:rg0.02h_m}
\end{minipage}
\end{figure}

In Fig. \ref{fig:rg0.02h_m} we display
the local magnetization density $\vec m$ of this CV
at points  on the $x$ axis, defined so that
 the phase difference
between the $u$ and $0$ components vanishes for $x > 0$.
Near the trap center $\vec m$ points mainly along the horizontal,
turning towards $\hat z$, the direction of
net magnetization, only further away.
The magnitude as well as the $z$-component of $\vec m$
depend only on the radial distance from the center of the
trap.
The azimuthal angle of $\vec m$ is
the negative of that of the corresponding physical point
in space. 
%(i.e., $\vec m$ forms a spiral)
It is interesting to note that the presence
of the CV may not be apparent if
one examines only the particle number density $n$, in 
strong contrast to the case of a scalar condensate.
\cite{Dodd97}. 
%edges ferro.

\begin{figure}[h]
\centerline
{ \epsfxsize=0.22\textwidth
\epsfysize=0.37\textwidth
\rotate[r]
{ \epsfbox{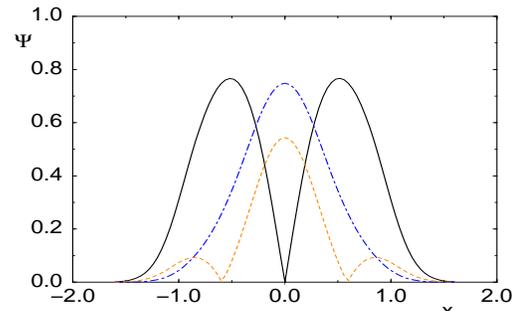} } }
\vskip 0.3 cm
\begin{minipage}{0.45\textwidth}
\caption[]{ CV for $\tilde c_2 = 0.02$, $\Omega = 0.45$,
and $m_{\rm tot} = 0.03 $ ($H=0 $).
 $|\Psi_u|$, dashed-dotted; $|\Psi_0|$, full line; $|\Psi_d|$, dashed.
}

%\vskip 0.3 cm
\label{fig:rg0.02d_psi}
\end{minipage}
\end{figure}

\begin{figure}[h]
\centerline
{ \epsfxsize=0.22\textwidth
\epsfysize=0.37\textwidth
\rotate[r]
{ \epsfbox{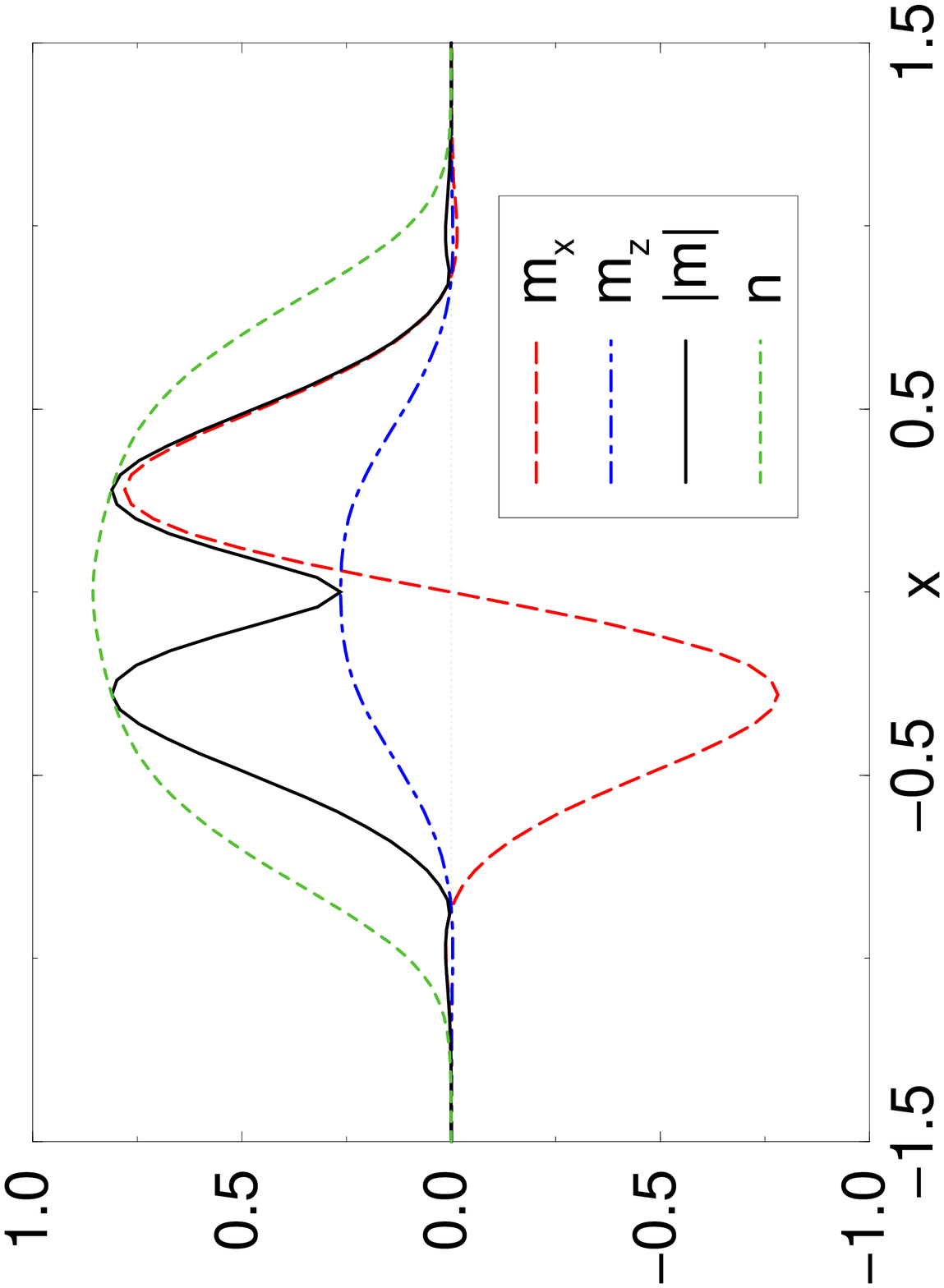} } }
\vskip 0.3 cm
\begin{minipage}{0.45\textwidth}
\caption[]{ Local magnetization
and density for the CV of Fig \ref{fig:rg0.02d_psi}.
 }

%\vskip 0.3 cm
\label{fig:rg0.02d_m}
\end{minipage}
\end{figure}

%%%%%%%%%%%%%%%%%%%%%%%%%%%%%%%%%%%%%%%%%%%%%%%%%%%%%
{\bf IV:} 
This is the most intriguing region.  At very small
$\tilde c_2$ (and not too small $\epsilon$'s)
 the CV has spontaneous (spin) symmetry breaking 
in the sense that it has  a net magnetization
even when $H = 0$. 
The configuration is similar to that of 
Fig. \ref{fig:rg0.2dz1} 
except now 
the numbers of  spin-up and spin-down
particles are no longer equal
(see Fig \ref{fig:rg0.02d_psi}).
This configuration is stable 
(i.e. the topology of the CV remains
the same except for a re-adjustment of the amplitudes
of $\Psi$'s)
so long as
the total magnetization is  close to that of the
`spontaneous' one. 
%Correspondingly
%the $\Psi_u$ vortices move out whereas the $\Psi_d$ vortices
%move in. 
% Since $\Psi_u \ne \Psi_d$ for general positions
%$m_y$ and $m_z$ are in general finite.  
The corresponding local magnetization density is
as shown in Fig \ref{fig:rg0.02d_m}.
The direction of $\vec m$ thus rotates from 
$-\hat x$ through $\hat z$ to $\hat x$ as one
moves along the physical $x$-axis.
The existence of this spontaneous $m_{\rm tot}$
means more of the local order parameter is axial
like, and thus this state is possible only for sufficiently
small $\tilde c_2$. 
%\cite{note}.
For larger $m_{\rm tot}$, 
this configuration gives way to that
of a $u$-vortex and a $d$ core (Fig \ref{fig:rg0.2hx2}),
which has a larger susceptibility, discussed earlier.

%%%%%%%%%%%%%%%%%%%%%%%%%%%%%%%%%%%%%%%%%%%%%%%%%%%%%%%%%%%kk

%%%%%%%%%%%%%%%%%%%%%%%%%%%%%%%%%%%%%%%%%%%%%%%%%%%%%%%%%%%%%

In conclusion I have shown that the internal vortex
structure of a spin-$1$ Bose condensate in a harmonic
trap is much richer than that of a condensate
with a scalar order parameter.  
%Observation of these
%objects would be interesting indeed.  
I thank T.-L. Ho for his comments on the manuscript.

%%%%%%%%%%%%%%%%%%%%%%%%%%%%%%%%%%%%%%%%%%%%%%%%%%%%%%%%%%%%

%%%%%%%%%%%%%%%%%%%%%%%%%%%%%%%%%%%%%%%%%%%%%%%%%%%%%%

%%%%%%%%%%%%%%%%%%%%%%%%%%%%%%%%%%%%%%%%%%%%%%%%%%%%%%%%%%%%%

%%%%%%%%%%%%%%%%%%%%%%%%%%%%%%%%%%%%%%%%%%%%%%%%%

%%%%%%%%%%%%%%%%%%%%%%%%%%%%%%%%%%%%%%%%%%%%%%%%%%%%%%%%%%%%%%%%%%%%%
%%%%%%%%%%%%%%%%%%%%%%%%%%%%%%%%%%%%%%%%%%%%%%%%%%%%%%%%%%%%%%%%%%%%%%%

%%%%%%%%%%%%%%%%%%%%%%
%%%%%%%%%%%%%%%%%%%%%%%%%%%%%%%%%%%%%%%%%%%%%%%%%%%%%%%%%%%%%
%%%%%%%%%%%%%%%%%%%%%%%%%%%%%%%%%%%%%%%%%%%%%%%%
%%%%%%%%%%%%%%%%%%%%%%%%%%%%%%%%%%%%%%%%%%%%%%%
%\begin{figure}[h]
%\centerline
%{ \epsfxsize=0.27\textwidth
%\epsfysize = 0.37\textwidth
%\rotate[r]
%{ \epsfbox{figs/rg0.2fc.ps} } }
%\vskip 0.3 cm
%\begin{minipage}{0.45\textwidth}
%\caption[]{ Normalized free energy $F(m)$ versus
%total magnetization for $\tilde c_2 = 0.2$, $\tilde \Omega = 0.45$
%(open symbols)
%and $\Omega = 0.5$ (filled symbols)
%. diamonds: Vortices consisting of
%vortices of $u$ and $d$, both circulation $1$,
%topology as that of Fig \ref{fig:rg0.2dx2} except
%adjustment of relative magnitude of $\Psi_u$ and $\Psi_d$;
%triangles right: vortex in $u$ but not $d$, topology
%as that of Fig \ref{fig:rg0.2hx2}.
%Also shown are the free energies for Vortices
%of the configuration as in Fig\ref{fig:rg0.2dz1}
%except $|\Psi_u| \ne |\Psi_d|$
%(triangles up).
%At $m_{\rm tot} = 0$ this configuration is
%equivalent to that represented by the diamonds.
% }

%\vskip 0.3 cm
%\label{fig:rg0.2fc}
%\end{minipage}
%\end{figure}

%%%%%%%%%%%%%%%%%%%%%%%%%%%%%%%%%%%%%%%%%%%%%%%%%%%%%%%%%%%
%%%%%%%%%%%%%%%%%%%%%%%%%%%%%%%%%%%%%%%%%%%%%%%%%%%%%%%%%%%%%

%%%%%%%%%%%%%%%%%%%%%%%%%%%%%%%%%%%%%%%%%%%%%%%%%%%%%%%%%%%%%
%%%%%%%%%%%%%%%%%%%%%%%%%%%%%%%%%%%%%%%%%%%%%%%%%%%%%
%%%%%%%%%%%%%%%%%%%%%%%%%%%%%%%%%%%%%%%%%%%%%%%%%%%%%%%%%%%%%
%%%%%%%%%%%%%%%%%%%%%%%%%%%%%%%%%%%%%%%%%%%%%%%%%%%%%%%%
%%%%%%%%%%%%%%%%%%%%%%%%%%%%%%%%%%%%%%%%%%%%%%%%%%%%%%%%%%%%%
%%%%%%%%%%%%%%%%%%%%%%%%%%%%%%%%%%%%%%%%%%%%%%%%%%%%%%%%%%%%%

%%%%%%%%%%%%%%%%%%%%%%%%%%%%%%%%%%%%%%%%%%%%%%%%%%%%%%%%%%%%%%%

%\end{multicols}


\begin{thebibliography}{10}
\bibitem{noteN} For comparison, for
 a cloud harmonically trapped in all three directions,
the number of particles in the TF approximation
is $ N_o^{(3)} = { R_z \over 15 a} ({R \over \lambda_0})^4 $
where $2 R_z$ is the dimension of the cloud in the $z$-direction.
\cite{Baym96}
For future reference for the parameters appropriate to $^{23}$Na,
 $N_o^{(3)} = 1.7 \times 10^4/ \Lambda$
if $\epsilon = (\lambda_o/R)^2 = 0.1$
for a trap with trap frequency $100$ Hz in the radial
direction. Here $\Lambda$ is the anisotropy parameter
$R/R_z$.
\end{thebibliography}
\end{document}